\documentclass{article}                     
\usepackage{graphics}
\usepackage{graphicx}
\usepackage{amsmath}
\usepackage{amsthm}
\usepackage{amsfonts}
\usepackage{amssymb}
\usepackage{url}
\usepackage{authblk}

\DeclareGraphicsExtensions{.pdf,.png,.jpg}
\input pdfcolor.tex 
\usepackage{graphicx}
\newtheorem{lemma}{Lemma}
\begin{document}

\title{Parallel Triangles Counting Using Pipelining}


  \author[1]{Juli\'an Ar\'aoz}     \author[2]{Cristina Zoltan}   \affil[1]{EIO Department} \affil[2]{Computer Science Department\\ Universitat Polit\`ecnica de Catalunya\\
 Barcelona, Spain}

\maketitle
\setcounter{tocdepth}{3}
\begin{abstract}
The generalized method to have a parallel solution to a computational problem, is to find a way to use \textit{Divide \& Conquer} paradigm in order to have processors acting on its own data and therefore all can be scheduled in parallel.  MapReduce is an example of this approach: Input data is transformed by the mappers, in order to feed  the reducers that can run in parallel.  In general this schema gives efficient problem solutions, but it stops being true when the replication factor grows. We present another program schema that is useful for describing problem solutions, that can exploit dynamic pipeline parallelism without having to deal with replication factors. We  present the schema in an example: counting triangles in graphs, in particular when the graph do not fit in memory. We describe the solution in NiMo, a graphical programming language that implements the \textit{implicitly functional parallel dataflow} model of computation.

The solution obtained using NiMo, is architecture agnostic and can be deployed in any parallel/distributed architecture adapting dynamically the processor usage to input characteristics.
\vspace{5 mm}

\end{abstract}

\section{Introduction}

With the emergence of commodity multicore architectures, exploiting
tightly-coupled parallelism has become increasingly important. Most of the parallelization efforts are addressed to applications that compute with large amounts of data in memory and in general have a regular behavior. 

Frameworks that implements MapReduce are a great success. A huge amount of papers show the willingness of using this  program schema to solve problems.  One can say that having a framewok, helps in testing the goodness of the solution but also because implementations deal with errors and the programmer need not take care of the deployment in parallel architectures is also done by the implementation.

In the scenario of tiny artifacts, interactive/reactive applications, big graphs, do not deal with  huge amounts of data in memory, but streams of data. But still  can exploit several cores and even distributed architectures.

The rest of the paper is organized as follows. In the
next section, we provide a brief background about the
typical strategy used to count triangles and a summary
of the previous work that is most related to our paper.
Section \ref{informal} provides an informal  description of our algorithm  using a Nimo implementation.
 Section \ref{formal} presents the algorithms and the proof of correctness.  
 We conclude
in Section \ref{CR}

\section{Problem History}
Very recently, a parallel algorithm for triangle counting using
MapReduce framework \cite{Suri:2011:CTC:1963405.1963491} has been proposed. Their algorithm is
exact and do not require to keep the entire input graph in memory
at each individual machine. We describe this algorithm in \ref{CTMapReduce}.

A divide an conquer approach to the problem is given in \cite{Arifuzzaman:2013:PPA:2505515.2505545}, using the BSP  model to synchronize the parallel workers and MPI to implement the algorithm. A collection of subgraphs,  of size equal to the number of processors,  of the given graph are constructed and due to the collection construction, triangles are counted by counting on each subgraph and accumulate the result.  The improvement over \cite{Suri:2011:CTC:1963405.1963491} is due to the
fact that their algorithm generates a huge volume of intermediate data, which are all possible 2-paths centered at each
node. Our improvement over \cite{Arifuzzaman:2013:PPA:2505515.2505545} is that we do not collect in a single machine all the adjacent nodes to a core node (responsible node), but only those that are not been collected by some other responsible node.   Also they need to do preconditioning in order to balance work load. Our dynamic scheduler is able to  balance work load based on the size of the neighbors of each responsible node. 

\section{Streaming Model of Computation} 

In \cite{McGregor:2014:GSA:2627692.2627694,survey}  we can find a good surveys on  algorithms  based on the streaming model of computation. 

In the Semi-Streaming model of computation, the input graph,
G = (V, E), is presented as a stream of edges (in any order),
and the storage space of an algorithm is bounded by $\Theta(n  \  polylog \ n)$. 

We are particularly interested in algorithms that use
only one pass over the input, but, for problems where this is provably
insufficient, we also look at algorithms using constant or, in some cases,
logarithmically many passes.

\section{MapReduce as a composition of two streaming algorithms}\label{CTMapReduce}

MapReduce can be described as a two face algorithm: first the set of input values are tansformed element by element into another set using the \textit{Map} operator,  preserving o not the cardinality of the set. In the second phase, this newly generated set is partitioned into equivalence classes  that are to be reduced using the \textit{Reduce} operation.  For example, for multiplying two matrices, the set of values $ (a_{i,k} * b_{k,j}, i,k,j)$ is generated in some order and then collect in the same class all the values with the same $i,j$.  Each equivalence class is reduced by adding all the elements in the class. In this case the communication cost is $n^3$.

\section{Pipeline program schema}
There are several sequential algorithms  for counting the number of triangles in a graph. If the graph is described as a matrix, there is a very simple algorithm:

$T = 0$ ;

for $v \in  V$ do

\ \ \ for $u \in \Gamma(v)$ do

\ \ \ \ \ for $w   \in \Gamma(v)$ do

\ \ \ \ \ \ \ if $((u;w) \in E)$ then

\ \ \ \ \ \ \ $T = T + 1/2$;

$return(T/3)$;

As seen in the algorithm each triangle is counted six times.

This algorithm assumes that the graph fits in memory, but for most problems the matrix describing the graph  is sparse or do not fit in memory.  
\subsection{Node Iterator in MapReduce}

In \cite{Suri:2011:CTC:1963405.1963491}  they present a composition of  two  MapReduce algorithms. In the application of MapReduce it geneartes  the set of 2-path, having a given responsible node.  He adds up the edges present in the graph and  the 2-path described by  their end nodes and cluster the elements using the pair of nodes as key. If both types of elements  are present in the same cluster, the number of triangles is equal the cluster size minus one. Otherwise the number of triangles in the cluster is zero. Adding the number of triangles in each cluster gives the total  number of triangles in the graph.  
\begin{itemize}
\item Round 1: Generate the possible  2-paths in
the graph.  The reduce step collects pairs of edges having a common node.
\item Round 2: Check which of the the 2-paths generated
in Round 1 can be closed by an edge in the graph
and count the triangles accordingly.
\end{itemize}

As all the MapReduce algorithms, a hashing function is used  to cluster elements that are to be treated by a single reducer.  It is usual that the number of reducers coincides with the number of available processors. So the behavior is not smooth   in he number of processors.  If the solution needs to be optimized, a more specific hashing function will be required, depending on the key structures. Algorithms having several rounds use the same hashing function no matter that the key domain can change.   In section \ref{informal} we will informally  present a pipeline algorithm for solving the triangle counting problem. We will use the  NiMo language to describe the algorithm because NiMo has the ability of dynamically generating programs and also because we can analyze the program execution  behavior as we will see in Sec. \ref{informal}. In Sect. \ref{language} a small language description is given in order to be able to follow the algorithm.

\section{The NiMo Language}\label{NiMo}\label{language}

NiMo (\textbf{N}ets \textbf{I}n \textbf{Mo}tion) is a graphic-functional-data flow language designed to visualize computer systems and their execution in an understandable way. This visualization (edition and/or execution) helps   to understand where and when resources are used in the program, thus giving clues to optimize  solutions. The language is very simple and is addressed to programmers (domain knowledge experts-DKE) and is  platform   independent.  The language is design to be simple: No variables and therefore no declarations. But strongly typed. 

The only element is the program, which is a net of  processes\footnote{A process need not to produce a result and most of the processes are functional.} communicating  using FIFO channels of unbounded capacity.  This boxes have input and output typed ports.

The DKE deals only with data and processes dependency, which is described using a data flow graph. This graph can have nodes having different degrees of granularity (Basic processes (atomic), processes that have a net as definition, black box processes).

There are no parallelization constructs in the language. At run time the schedulers  are in charge of assigning  processors to processes. The execution model is an asynchronous version of the BSP model\cite{Valiant:1990:BMP:79173.79181} allowing communication and execution taking place at the same time without race conditions. 

For testing algorithms, there is a IDE called NiMoToons, that assumes an unbounded number of processors.

The system gives the  function plot showing the available parallelism at each execution step    in fine
grain parallelism.  This gives an upper bound on obtainable parallelism for the given solution.  

Once certified the solution, if necessary, an expert on code parallelization can tune the code  to assist schedulers in their task of assigning processes to processors.   The solution need  not to be  changed, only tunned by defining an scheduler for it.



As the language inherits from functional programming, and processes are first class citizens (can be manipulated, created and destroyed as any other data element), problems with  low natural parallel (imperative or functional) solutions, can be solved in different approaches, using dynamic data structures.
This allows a systematic way of generating dynamically communication paths. 

Another characteristic derived from the combination of data flow and functional programming is the ability to do concurrently processing and communication using pipeline parallelism.

The characteristic of the language is that allows to see and analyze data flow in a system at different levels of granularity and also as a workbench for designing high tunned problem solutions.

Another characteristic is that the programs are deterministic unless non-deterministic constructs are used \cite{Lee:EECS-2006-1,Bocchino:2009}.  A NiMo programmer need  not to be aware of the number of processor that will be used in running the program. A dynamic scheduler for executing transformations on a NiMo program  scales up to the number of available processors.

Rather than a series of imperative commands sequentially executed, a dataflow program is more like a network of workers in an assembly line who do their work to the extent that they have the necessary material. But unlike a worker, processes can receive several inputs  and provide several outputs.   In particular, workers in  NiMo have different attitudes toward work, allowing a better division of labor. A worker can awake other workers in order to complete his task. 

 NiMo, having very few constructs,  is not a visual interpreter for a textual language because there is no textual code at all. It is a true graphical programming language.  Programs are process networks  (PN) according to \cite{CH-book} 
 or operator nets as in \cite{springerlink:10.1007/BFb0027039}. Unlike operator nets, NiMo nets deal with partial operators, allowing processes to be blocked on erroneous inputs. The program exploits the bi-dimensional description  to express data dependency and color to describe modes and states. The two-dimensional display eliminates the need for operator precedence. 
Processes can be specialized by parameters on the top. Processes communicate  through unbound capacity  channels, acting as FIFO queues that  carry references any type of value from basic NiMo types: integers, reals, booleans,  processes. But also references to structured types as lists or tuples.

FIFO queues can also carry references to  expressions, implementing \textit{futures}.   It has  additional characteristics as wait by necessity, also called blocking read, no explicit PUT or GET operations and value sharing.
  Programs in execution fully exhibit their state. Ready processes act concurrently assuring no data races.  
  
  NiMo programs are strong typed,  but the language does not have type declarations. Processes may be  polymorphic, higher oder and have multiple outputs. The language has a set of primitive processes well suited for stream processing and supports  executing  programs and interactive debugging. The system provides an also graphical and incremental type inference system that guarantees program type-safeness by construction. A more complete description can be found in \cite{art-revista}. Some program examples can be seen in \cite{nimo-home}.  Fine grain  concurrency is supported by independent computation and  by the  language construct: pipelining. NiMo systems can be deployed in a distributed architecture.

We will use the notation used in \cite{galois}, because although Galois is an extension of $C^{++}$, shares a set of concepts with NiMo\footnote{CnC also has the notion of a \textit{step} that can execute-meaning that in a state   if the process  finds a processor, it is able to execute}.

 One measure of amorphous data-parallelism is
the number of active nodes (fireable processes)  that can be processed in parallel at each
step of the algorithm for a given input, assuming that:
\begin{itemize}
\item  there is an unbounded number of processors, 
\item  an activity takes one time step to execute, 
\item  the system has perfect knowledge of neighborhood and ordering constraints so it only executes activities that can
complete successfully, and 
\item a maximal set of activities, subject to neighborhood and ordering constraints, is executed at each step.
\end{itemize}
This is called the available parallelism at each step, and a function plot showing the available parallelism at each step of execution of
an irregular algorithm for a given input is called a \textit{parallelism profile}.

\subsection{Informal Algorithm presentation}\label{informal}

We assume the input is a large graph given by the enumeration of its edges. We assume a non oriented graph and there are no duplicated edges\footnote{Oriented edges are a special case of the one treated here}.

The general idea of the  two round schema we use is identical to the node-iterator schema used by  \cite{Suri:2011:CTC:1963405.1963491}. The main difference is the way the possible 2-two path are identified and the program structure.

We use the same principle of a single node is  ``responsible"  for making sure the triangle gets counted. In  \cite{citeulike:8568862}   this is obtained via the knowledge of the degree of each node. Dealing with graphs large enough that do not fit in memory, this additional knowledge requires an additional traversal on the edges of the graph which is not needed in our approach.

The algorithm we present is implemented in NiMo as a sequence of actors (processes) that will change their role (mutate its behavior) when  enough knowledge has been collected.  Our actors first role  is to acquire an edge and become a process that is ``responsible'' for this first node in the edge. The responsible actor receiving an edge will collect in his memory all the nodes, not being collected by another responsible, adjacent  to the``responsible'' node. Other edges, not adjacent to the responsible node,  are passed to his neighbor.  When there are no more edges in the first input,  each actor changes again his role, receiving again the sequence of edges and counting a triangle, whenever the edge forms a triangle with  the ``responsible ''  node and two adjacent ones.  Whenever all the edges are analyzed, the process passes its triangle count to its neighbor and dies.

We will use a small graph to show the algorithm behavior in this case.
 \begin{figure}[h]
	\centering
	\includegraphics[width=\textwidth]{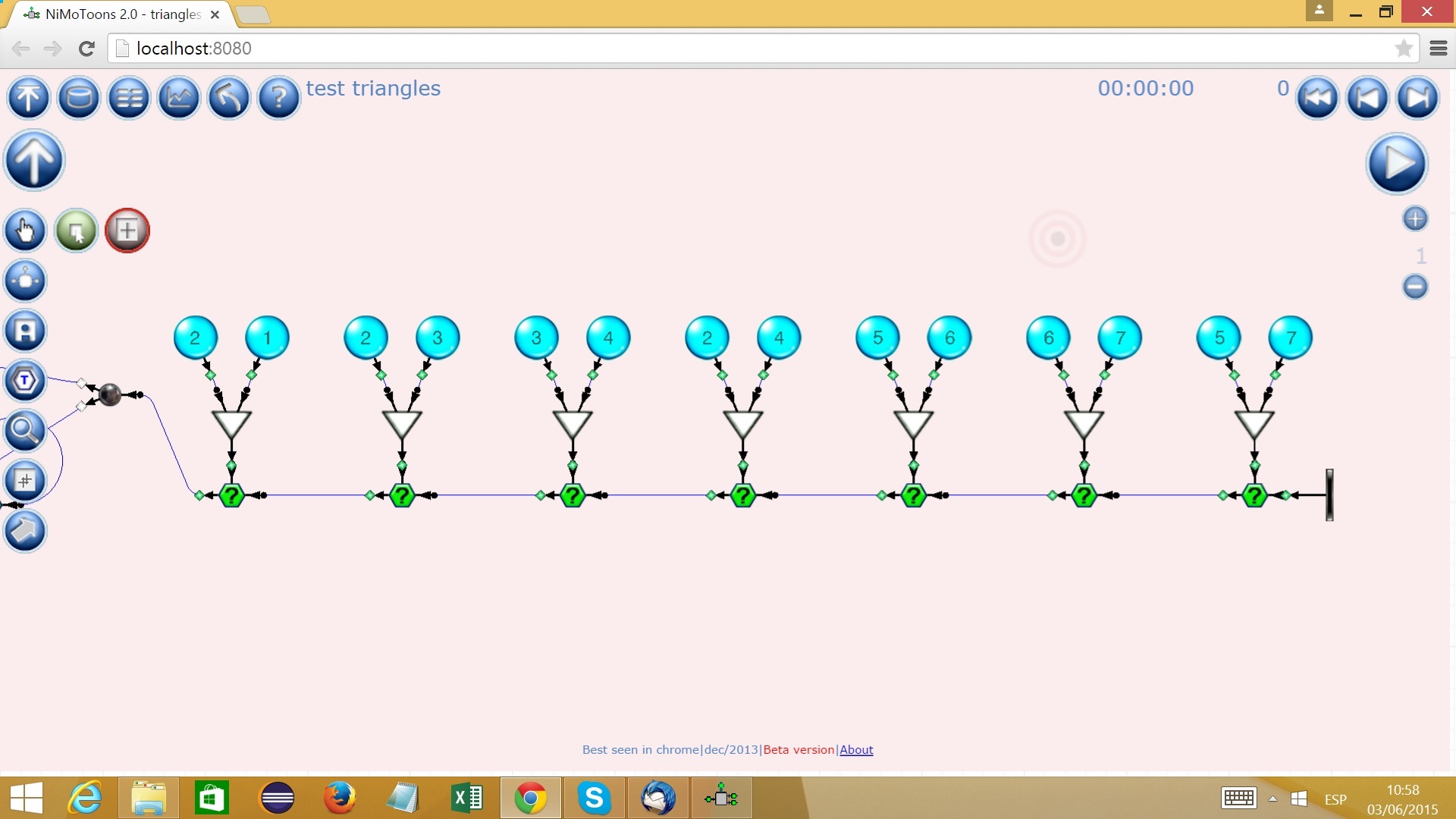}
	\caption{Graph represented by a list of edges i.e. 2-node tuples}
	\label{data}
\end{figure}

 \begin{figure}[h]
	\centering
	\includegraphics[width=\textwidth]{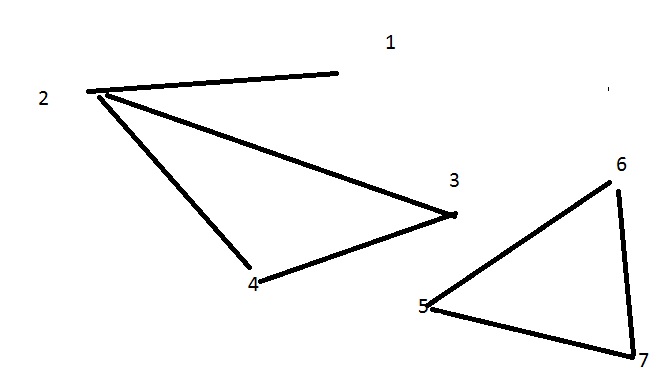}
	\caption{Graph}
	\label{graph}
\end{figure}

 \begin{figure}[h]
	\centering
	\includegraphics[width=\textwidth]{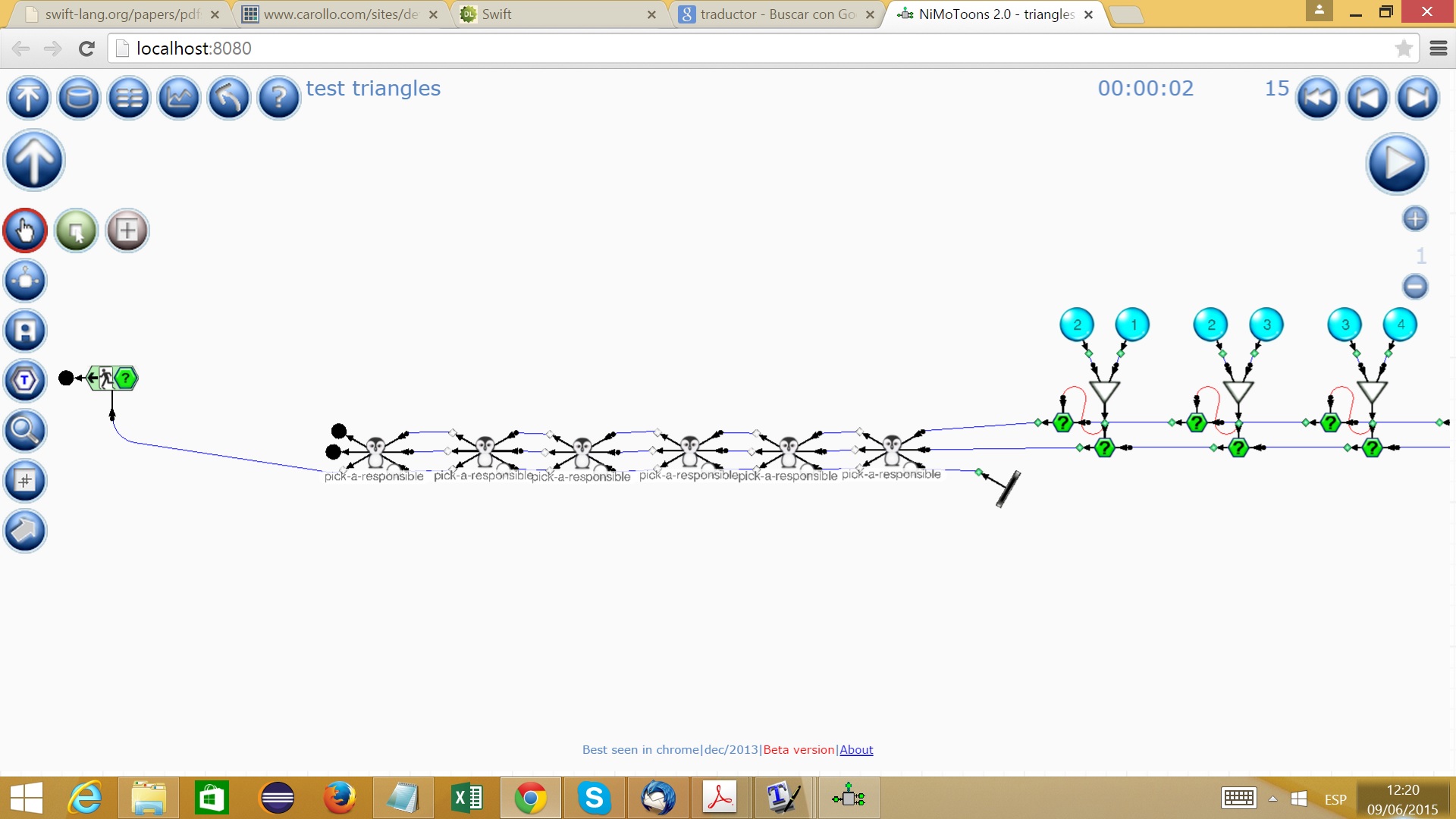}
	\caption{Pick a responsible sequence}
	\label{Pick}
\end{figure}

In Fig. \ref{data} we see the NiMo graph representation.  Is represented as a sequence of pairs (2-tuples) ending by an end of sequence symbol : \includegraphics[width=0.2 in]{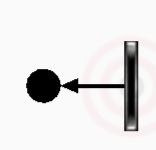}.  The sequence represents the graph in Fig. \ref{graph}.

The algorithm starts with a program that generates a composition of $|V|-1$ actors/processes, called \textit{pick-a-responsible}, having the same role. They are represented by penguins. The process name and icon are chosen arbitrarily. They  shake 3 hands with each neighbor as can be see in Fig. \ref{Pick}. The graph, described as a sequence of edges,  flows through the first hand. The algorithm will be described in two rounds: each round reads the sequence of edges describing the graph.

\begin{figure}[h]
	\centering
	\includegraphics[width=\textwidth]{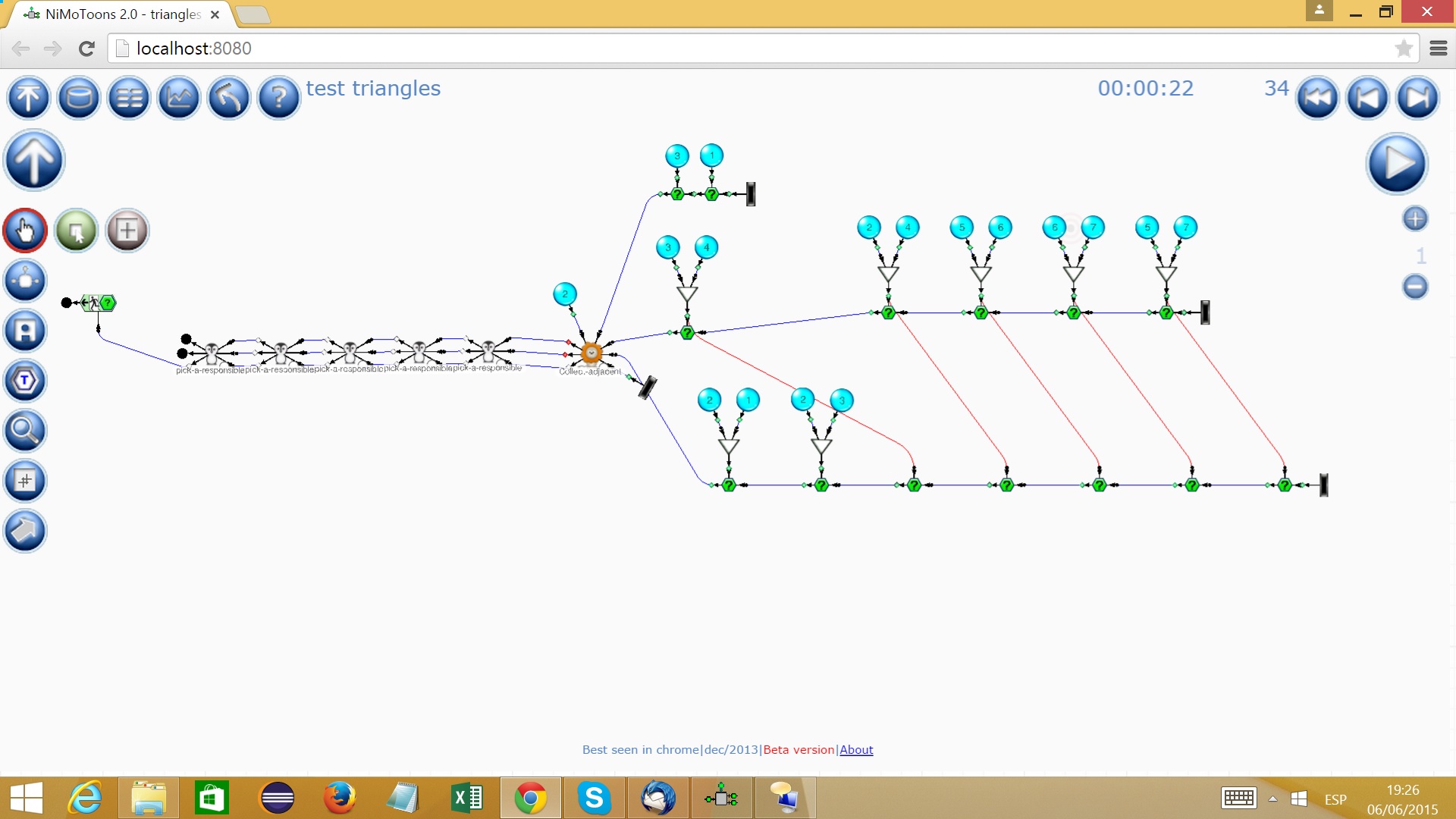}
	\caption{All the actors are set in place and  one has found two adjacent edges}
	\label{stage0.1}
\end{figure}
\begin{itemize}
\item Round 1:  Whenever \textit{pick-a-responsible}/penguin receives an edge   $(a,b)$ changes its role becoming  \textit{Collect-adjacent}/lion process having as parameters a  responsible node  $a$ and a list of adjacent nodes, with only one element $b$.  \textit{Collect-adjacent} processes are represented by lions. The icon used for processes serves to find on the screen when several instances of the same process is present during a computation and easily identifying processes without reading its name.

In figure Fig \ref{stage0.1} we see the program execution, where  \textit{Collect-adjacent} has collected 2 adjacent node for the responsible node 2.

\begin{figure}[h]
	\centering
	\includegraphics[width=\textwidth]{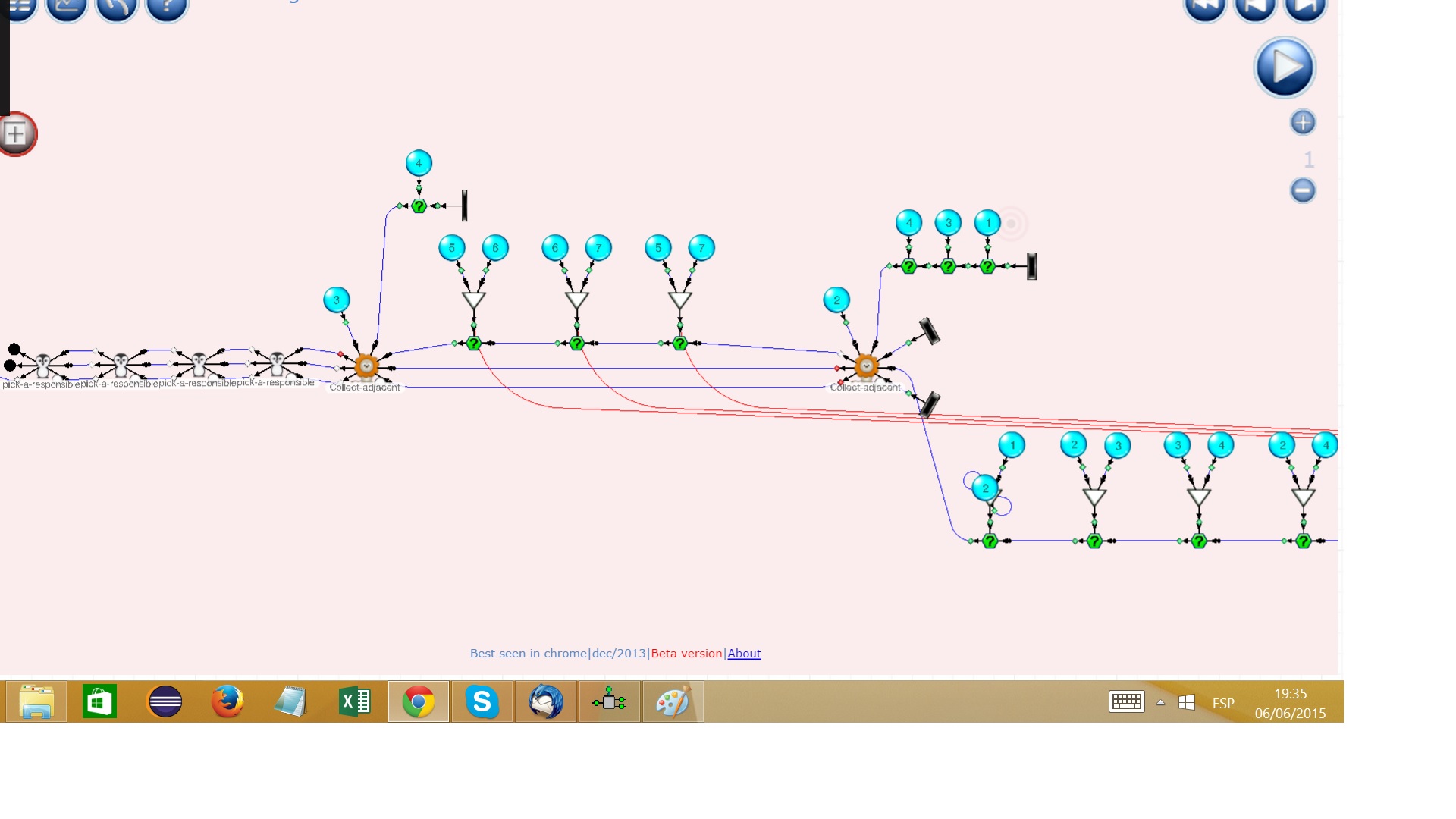}
	\caption{There are two responsibles and some adjacent}
	\label{stage1}
\end{figure}
If there are no more edges, \textit{pick-a-responsible}/penguin exits\footnote{Disconnects its outputs and inputs and reconnect its inputs with its outputs i.e. becomes an identity process. If an actor will not produce results anymore, can die and will be erased by the garbage-collector}.

Therefore  \textit{pick-a-responsible}/penguin is present only in the first round and is changed by another process or fades away whenever a single input shows in its input parameter.

The role of  \textit{Collect-adjacent}/lion, having as responsible node $a$,   is to receive the edges, one after the other and when receiving an edge $(c,d)$, pass it to the left neighbor if  $ c \neq a\ \wedge d \neq a$. Otherwise it adds to the set of adjacent nodes the node which is not equal to $a$.

In figure Fig. \ref{stage1}   the \textit{Collect-adjacent}/lion process  for responsible node 2 is to receive an end. The \textit{Collect-adjacent}/lion process  for responsible node 3 has collected only one adjacent node.

When  \textit{Collect-adjacent}/lion  receives an end i.e. there are no more edges, it mutates its behavior into the role of \textit{Count-triangles}/toucan and sends an end to his neighbor  using his firsthand. Before mutating  \textit{Collect-adjacent}/lion  has collected a subset of the edges adjacent to the responsible node.

\begin{figure}[h]
	\centering
	\includegraphics[width=\textwidth]{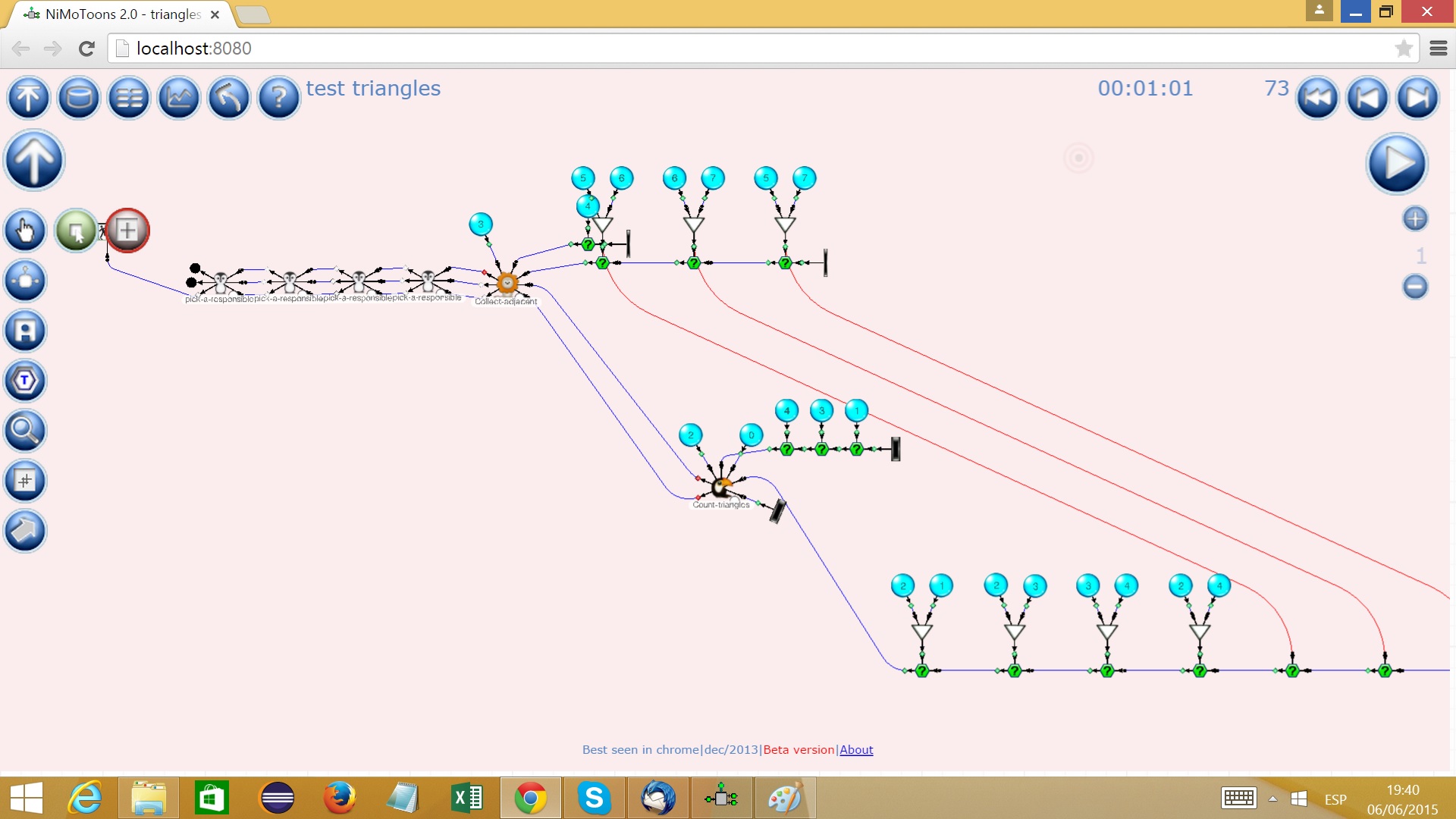}
	\caption{Start to check edges for closing triangles}
	\label{stage2}
\end{figure}
\textit{Count-triangles}/toucan will keep the responsible node and all its adjacent ones collected by the corresponding  \textit{Collect-adjacent}/lion  actor and record that no triangles showed up so far as shown in Fig. \ref{stage2} having a third parameter set to $0$. The actors  will have only two hands to shake with its neighbors: one hand to deal with the graph and the second hand to propagate the local triangle count.  

In figure Fig: \ref{stage2} we see the process \textit{Count-triangles}/toucan  prepared to count triangles that include the resposible node (2).

\item Round 2: \textit{Count-triangles}/toucan  receives an edge and checks  if  both ends are adjacent to the responsible node. If so, it adds one to number of triangles found. In any case the edge is passed to its left neighbor (produced on its output), using its first hand. 

If  \textit{Count-triangles}/toucan receives the signal that there are no more edges, it passes on its first hand the signal that there are no more edges and it passes its count to the left neighbor added to the input in his second hand, using its second hand and dies\footnote{An actor can die connecting its outputs to its inputs or simply  disconnecting  its outputs. If an actor will not produce results anymore, can exit the scene-will be erased by the garbage-collector}.
\end{itemize}

 \begin{figure}[h]
	\centering
	\includegraphics[width=\textwidth]{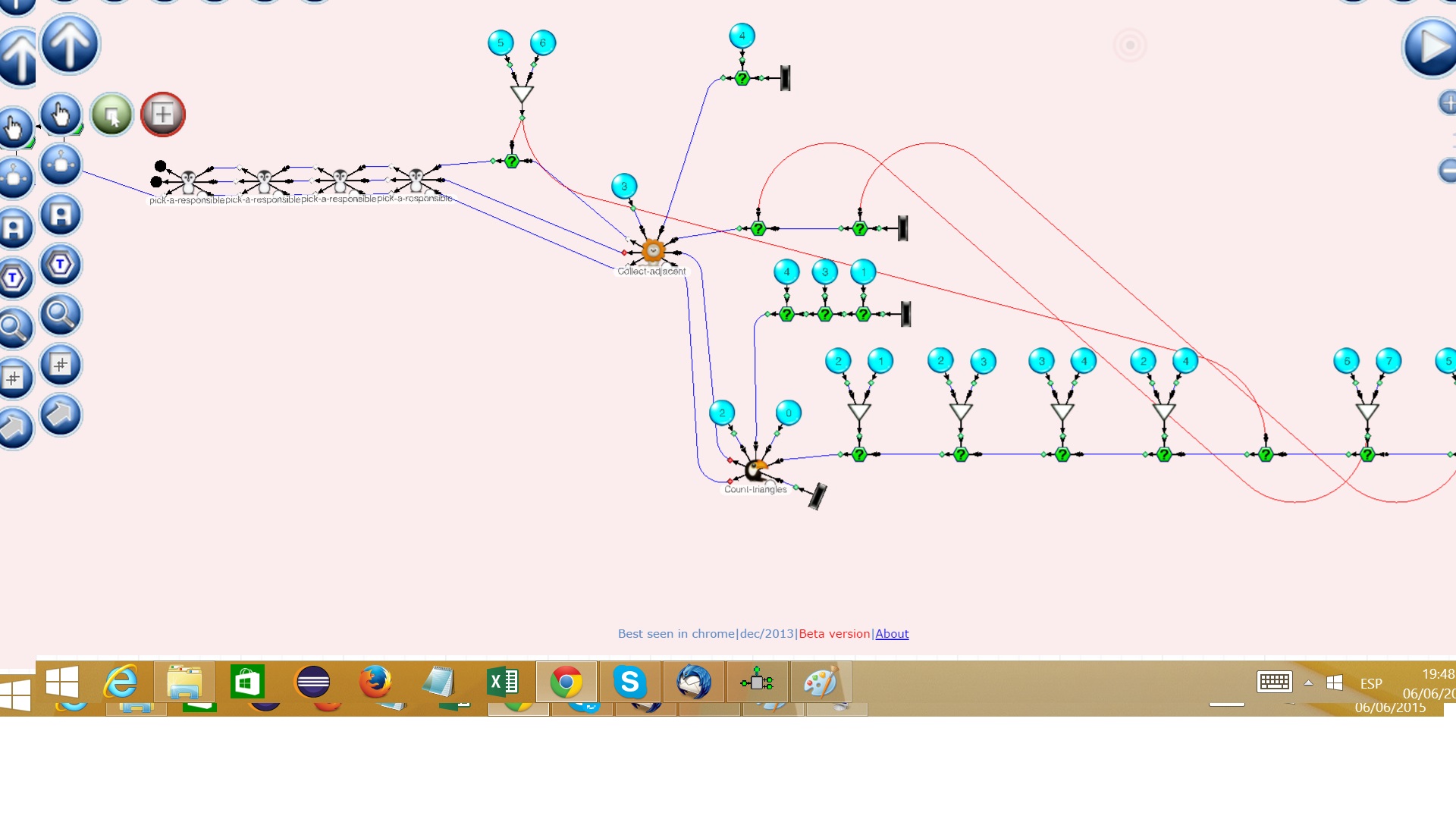}
	\caption{Fives  will become responsible}
	\label{stage0.0}
\end{figure}

	In Fig. \ref{stage0.0} we see that the penguin is about to transform node  5 into a responsible node. As we can see there are three diffent animals in the picture, showing that NiMo evaluation model, implements BSP without barriers i.e. communication and execution can take place at the same time.

\begin{figure}[h]
	\centering
	\includegraphics[width=\textwidth]{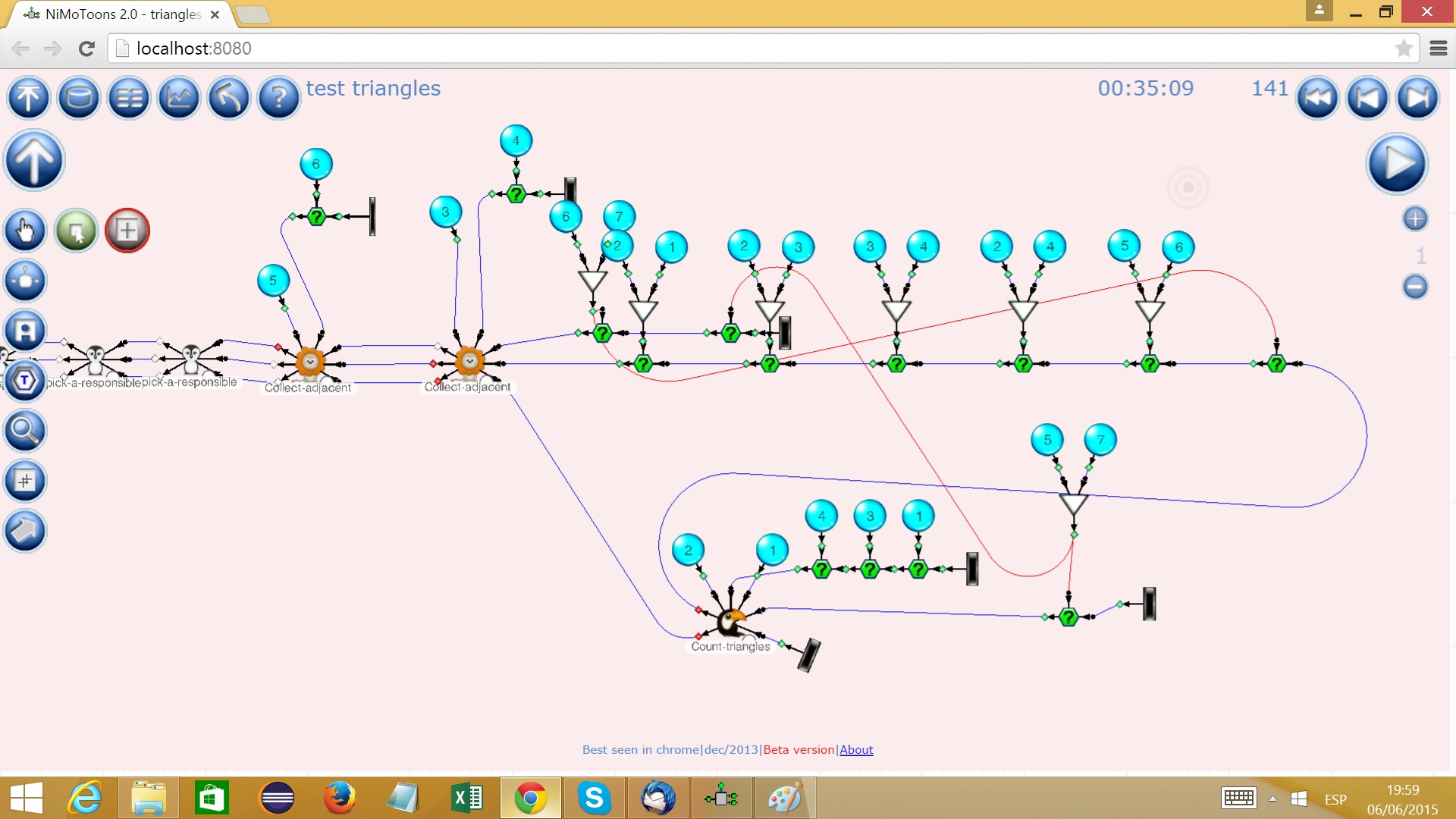}
	\caption{A  triangle found}
	\label{stage2.1}
\end{figure}

In Fig. \ref{stage2.1} the toucan has already found a triangle.

	\section{Formalizing the Algorithm}		
\subsection{Notation}

The given graph  is denoted by $G(V,E)$, where $V$ and $E$
are the sets of vertices (nodes) and edges, respectively, with
$m = |E|$ edges and $n = |V |$ vertices labeled, label that can be compared by equality. We use the words node and vertex interchangeably. We
assume that the input graph is undirected, without loops or multiple  edges. If $(u, v) \in E$,
we say $u$ and $v$ are neighbors of each other. The set of
all neighbors of $v \in V$ is denoted by $N_v$, i.e., $N_v = \{u \in V |(u, v) \in E\}$.
A triangle is a set of three nodes $u, v,w \in  V$ such that
there is an edge between each pair of these three nodes, i.e.,
$(u, v), (v,w), (w, u) \in E$. The number of triangles containing
node v (in other words, triangles incident on v) is denoted
by $T_v$. Notice that the number of triangles containing node
$v$ is as same as the number of edges among the neighbors of
$v$, i.e., $T_v = | \{(u,w) \in  E | u,w \in N_v\} |$. 

\subsection{Algorithm}\label{formal}

The algorithm for counting triangles is design using two rounds i.e. reading twice the graph. The first round collects adjacent nodes to a ``responsible'' node that can become part of a triangle that includes the responsible node.

NiMo implements bounded and unbounded minimization \cite{opac-b1083142} as a program schema. This schema is generalized to allow processes having  multiple outputs. 
 This schema, having  sequences as input and the process being composed can be applied element by element, allows to describe computations that  can exploit pipeline parallelism.

The program is implemented as a fixed point solution of the composition of \textit{pick-a-responsible}/penguin processes that will produce a sequence of integers counting the number of tiangles found  fo each ``responsible'' node.

\textit{pick-a-responsible}/penguin  is the process obtained from the \textit{combination}, denoted by $\times$, of three processes: the first acting on the first port and the other two the blocking process\footnote{Really, NiMo processes will not use input from the second and third pots until changing  role.} acting on the second and third inputs.

     \[\text{\textit{pick-a-responsible}} (x,y,z) = (F_1 \times F \times F)(x,y,z)\]
  
     \textit{collectadjacent}/lion is a pocess obtained as the combination of three process as    \textit{pick-a-responsible}.
     
       \[\text{\textit{collectadjacent}} (x,y,z) = (F_2\times F \times F)(x,y,z)\]
       
  \textit{Count-triangles} is a process, obtained from the transformation of      \textit{collectadjacent}, loosing its first input. The second input serves only to tie together the individual results.
  
         \[\text{\textit{Count-triangles}} (x,y) = (F_3 \times F)(x,y)\]

The auxiliary function $F$ is used to simulate a function that do not consume its input and is used to describe that  \textit{pick-a-responsible}/penguin does not consume inputs on the second and third inputs.  Same happens with \textit{Collectadjacent} process.

\[\text{\textit{F}}(c) =
  \begin{cases}
    F(c)  & \quad \text{if } c = (a,b) : m\\
    eof & \quad \text{if }   c= eof\\
  \end{cases}
\]

$$\text{\textit{F}}_1(c) =
  \begin{cases}
    F_2 (a,[b]) (m)  & \quad \text{if } c = (a,b) : m\\
    eof & \quad \text{if }   c= eof\\
  \end{cases}
$$

\[F_2(r,ad) \times F \times F (c, x,y) =
  \begin{cases}
  ( F_2 (r,[s:ad]) \times F \times F)(u,x,y)  & \quad \text{if } c = (m,n) : u\and \\
                                        &  \quad( m=r  \lor n =r)  \ s \neq r\\
                                        & \quad (s = m \lor s =n)\\
    (c : F_2 (r,ad)  \times F \times F) (u,x,y)  & \quad \text{if } c = (m,n): u \and  \\
                                                &\quad( m\neq r  \land  n \neq r) \\
 (F_3(r,ad,0)  \times F )(x,y)& \quad \text{if }   c= eof\\
  \end{cases}
\]

\[(F_3(r,ad,i) \times F) (c,d) =
  \begin{cases}
   (c:\text F_3(r,ad,i)  \times F)(u,d)& \quad \text{if } c = (a,b): u\land \\ 
                                                 &\quad ( a \notin ad \land b \notin ad) \\
    (c:\text F_3(r,ad,i +1)  \times F) (u,d) & \quad \text{if } c = (a,b) \land  \\
                                                         & \quad( a \in ad \land b \in ad) \\
 \pi \times i & \quad \text{if }   c= eof\\
  \end{cases}
\]
In Round 1, the program is a sequence/composition  of  \textit{pick-a-responsible} processes transformed, due to the information collected from the incoming edges   into  a sequence of   \textit{Collect-adjacent} processes, with edges flowing on the first hand.

\begin{lemma}  The leftmost  \textit{Collect-adjacent} process has the first output hand not bound. No edge will arrive to this port.
\end{lemma} 
\begin{proof}
The initial program is a sequence   of  \textit{pick-a-responsible} processes  has length $|V|-1$.  The total number of responsible nodes is bound by  $||-1$. Therefore, every edge will be consumed by either a \textit{pick-a-responsible} process  or by a \textit{Collect-adjacent} process  therefore no edge will show at the leftmost process. 

\end{proof}

\begin{lemma} Every edge in the graph is encoded by  exactly  one \textit{Collect-adjacent} process.
\end{lemma} 
\begin{proof}
If an edge $(a,b)$ shows up in the first input of  a \textit{pick-a-responsible} process, it means that  neither $a$ nor $b$ where chosen as responsible nodes yet. So $a$ becomes a responsible node and the edge is recorded in the corresponding \textit{Collect-adjacent}  process and not passed to the neighbor. The only thing to prove is that there are enough \textit{pick-a-responsible} processes to become  \textit{Collect-adjacent}  processes. But the number of possible  \textit{pick-a-responsible}  processes is bounded by $|V|-1$ by construction\footnote{This bound is attained by a complete graph}.

If an edge $(a,b)$ shows up in the first input of  a \textit{Collect-adjacent} processes having as responsible $a$ or $b$, will be encoded by the process. Otherwise the edge will be passes to the neighbor process.
\end{proof}
\begin{lemma}When starting to count triangles, every  2-path that forms a triangle (closed wedge according to \cite{DBLP:journals/corr/JhaSP13}) is encoded in exactly one filter: the actor that holds it has a responsible node  which  is the path middle node.
\end{lemma}
\begin{proof}
If there is a triangle in the graph we need to show that one and only one of its path of length two is recorded in some filter.

Each filter $(a, a_1,a_2,\hdots , a_n)$ encodes  paths  $(a_i,a,a_j)$ with $i \ne j$. 

If there is a triangle, it includes three  2-paths. As edges are consumed by  filters, edges are present in at most one filter.
If there is a triangle that includes nodes $(a,b,c)$, lets assume that the first edge of the triangle that shows up in the enumeration of edges is the edge $(a,b)$. This edge is recorded  by some process having $a$ or $b$ as responsible. Let´s assume $a$ is the responsible node.  There are four possible subsequences that represent the given triangle:

\begin{enumerate}
\item $(b,c)\   (a,c)$ 
\item $(b,c) \  (c,a)$
\item $(c,b) \ (a,c)$
\item  $(c,b) \ (c,a)$
\end{enumerate} 
In any case, edge $(b,c) or (c,b)$ is sent to other filter and this filter keeps the edge ($(a,c)$. So no more 2-path  of the triangle can be retained by any other filter because  two edges are already retained by the filter that has $a$ as responsible.
\end{proof}

In Round 2, data flows in the first hand, until all the input is consumed and then the second hand collects all the locally counted triangles.

The final result is a sequences of integers, each integer counting the number of triangles including a responsible node. Adding up this sequence gives the desired result.
\section{What happens if the graph is not a simple one?}

In \cite{DBLP:journals/corr/JhaSP13} signal out the problem of estimating triangles when the graph is not simple. They show that approximate algorithms does not behave the same as the graph is not simple and its size increases.  They mention that for eliminating duplicated edges requires an additional  pass over the edges. Due to the structure of our algorithm, duplicates can be eliminated by replacing the operation of prefixing the newly arrived node in the  \textit{collectadjacent} processes by an union operation on sets.  

If instead, we want to count the number of triangles, even if they a repeated ones, instead of adding to a set, the operation will be adding the newly arrived to the multiset i.e. the set remembers the number of times the edge has appeared. When counting triangles, a closing edge, closes as many triangles as the minimum of the multiplicity of their endpoints. The algorithm  modularity allows to replace processes to easily adapt the algorithm  to other input characteristics.

The algorithm assumes that there is enough memory in each processor to hold the nodes adjacent to the responsible node. If it is not the case, this set can be stored in another memory.  Also, if necessary to speed up the process, this adjacent nodes can be stored in a hash table. This change amounts to change the \textit{close?} process to verify the condition on a pair of hash tables, instead of operating with a pair of node lists.
 In case of error, a similar schema can be followed.  Changing channels by processes that can retry reading in case of processors unable to complete the processing of a particular edge.

\section{Related work}

Introduced by Gilles Kahn, the Kahn Process Networks approached this
problem by having sequential processes (nodes) to communicate via unbounded
FIFO queues as message passing protocol\cite{Kah74}. Difference between Kahn Processes Networks and NiMo programs, is that NiMo processes need not to be sequential. NiMo processes have different granularities: Black boxes written in another language or NiMo nets written in NiMo.

In \cite{Wilde:2011:SLD:2286659.2286714} the Swift/T language is described. Is a language implementing the 
“implicitly parallel functional dataflow” (IPFD) model centered mainly in being able to have as input a variety of data sources. The programming model is very simple and forces the user to think on in-memory data. The language has the ability of integrating  leaf tasks written in other languages. Using this feature has proven very effective in implementing solutions in several domains. But the simplicity of the language does not allow some solutions that fully exploit architectures with a massive number of processors. The promote Many-Task Computing.
Swift does 4 important things for you:
\begin{enumerate}
			\item 	Makes	parallelism	more	transparent
		
				implicitly	parallel	functional	dataflow	programming	
\item Makes	computing	location 	more	transparent	

		runs	your	script	on 	multiple	distributed	sites	and	
diverse	computing	resources	(desktop	to	petascale)
\item Makes	basic	 failure	recovery	transparent	
			Retries/relocates	failing	tasks	
			Can	restart	failing	runs	from	point	of	failure	
\item Records	provenance	of	data	derivation	
			Made	possible	through	functional	encapsulation	
			
\end{enumerate}			

		The  mechanism  used is the instruction forall  that spawns processes for each body in the iteration.  In dependent iterations it does not give a high level of parallelism. NiMo has the \textit{Map} process and  recursion as mechanisms for spawning new processes required by the program execution.
\section{Concluding Remarks}\label{CR}

		We have used counting triangles  problem on a graph described as a sequence of edges as a NiMo programming example.   In the example we  see that a big amount of processors can be exploited, but the algorithm is correct even in a single processor semantics. The solution exhibits NiMo characteristics as the generation of a pipe of processes that simulate the well known bucket sorting method that separates the input set into disjoint classes. 	Processes in the pipe, change behavior as soon as enough data has been consumed. NiMo uses a BSP model of computation without barriers.  Therefore the algorithm presented here has in some sense  wavefront behavior when executed as BSP without barriers. 	

\end{document}